\documentclass[a4paper]{jpconf}
\usepackage{graphicx}
\usepackage{amssymb, amsmath} 
\usepackage{cancel} 
\usepackage{cite}

\newcommand{\be}{\begin{equation}}
\newcommand{\ee}{\end{equation}}
\newcommand{\bary}{\begin{eqnarray}}
\newcommand{\eary}{\end{eqnarray}}

\begin{document}

\title{Study of PeV neutrinos around dwarf galaxies near giant lobes of Centaurus A}
\author{E. Aguilar-Ruiz$^1$, N. Fraija$^1$, A.~Galv\'an-G\'amez$^1$, J. A. De Diego$^1$  and  A.~Marinelli$^2$ }

\address{$^1$ Instituto de Astronom\' ia, Universidad Nacional Aut\'onoma de M\'exico, Circuito Exterior, C.U., A. Postal 70-264, 04510 M\'exico City, M\'exico\\
 $^2$ I.N.F.N. \& Physics Institute Polo Fibonacci Largo B. Pontecorvo, 3 - 56127 Pisa, Italy}

\ead{eaguilar@astro.unam.mx, nifraija@astro.unam.mx, agalvan@astro.unam.mx, jdo@astro.unam.mx  and antonio.marinelli@pi.infn.it}

\begin{abstract}
The origin of recently discovered PeV neutrinos is an unsolved problem. In this work we consider a hadronic scenario to produce PeV neutrinos from a region around giant lobes of Centaurus A.   Although ultrahigh-energy cosmic rays (UHECRs) are accelerated and confined by giant lobes,  they can escape to be later injected in the inter-group medium where galaxies near the giant lobes provides the condition to confine them.  UHECRs interact with low-energy photons and protons producing high-energy photons and neutrinos. We found that the IC35 event cannot be generated neither inside the giant lobes nor galaxies close to the lobes of Centaurus A.
\end{abstract}

\section{Introduction}
Neutrinos are considered perfect astronomical messenger due to their very low capacity to interact with matter and radiation.  They are able to escape from environment in which they are created and because of the lack of electric charge, they can travel large distances without being deflected by magnetic fields.  Therefore, high-energy neutrinos could give us an indirect signal of the cosmic ray origins  which have been a mystery up to now.   Recently, the neutrino observatory IceCube reported the detection of high-energy neutrinos with extraterrestrial origin in the HESE (High Energy Starting Events) Catalog \cite{IceCube2013}, but no event has been associated with some known astrophysical source.    Multiples sources of neutrinos have been proposed like Gamma Ray Burst (GRB) \cite{Waxmand&Bahcall1997, Mezaros2001, Musare&Ioka2013, Fraija2014a, Fraija2016a}, Active Galactic Nuclei (AGN) \cite{Cholis2013, Fraija2014b, Fraija2016b} and Star-Burst Galaxies (SBGs) \cite{Loeb2006, Chang2015}.\\
On the other hand,  Pierre Auger Observatory (PAO) reported the detection of more than 100 ultrahigh-energy cosmic ray (UHECR) events \cite{PAO2007}. At these energies, UHECRs are attenuated by Cosmic Microwave Background (CMB). Therefore, sources of UHECRs are expected in distances around $\lesssim$ 75 - 100 Mpc if the composition are predominantly protons or Fe nuclei and closer if the predominant composition is lighter nuclei, as reported by PAO \cite{PAO2014}. One the best candidates is Centaurus A, classified like a FR-I radio galaxy.   It is the nearest active galactic nuclei to the Earth and the brightest source in radio-band on the sky (see for more detailed, \cite{Israel1998}).  This source has been proposed as a natural candidate to accelerate UHECRs \cite{Dermer2009,Fraija2012}. Moreover, PAO reported UHECRs over density around Cen A \cite{Aab2015}, although the acceleration mechanism  is not clear yet. In addition,  Cen A is inside the angular error of the most energetic neutrino ever detected IC35 which was reported by IceCube with an energy 2004$^{+236}_{-263}$ TeV .
\section{Neutrinos and Cosmic Rays}
To know if a source is able to accelerate cosmic rays  up to a certain energy we use the Hillas criterium that establish the maximum energy that CRs can reach on a source of size (R) and the strength of  magnetic field (B) \cite{Hillas1984}
\be
E_{\rm Z, max}= 17 \,  {\rm EeV}\, \, \rm Z \,\left(\frac{B}{\rm 1\,\mu G}\right)\,\left(\frac{R}{\rm 100\, kpc}\right)\,\left(\frac{\beta}{0.2}\right)\,.
\ee
Here,  Z is the atomic number and $\beta$ is the shock velocity 
The acceleration  timescale  $t_{\rm acc}\simeq \eta R_g$ with    $R_g=\frac{E_{\rm}}{eB}$ is given by
{\small
	\be\label{tac}
	t_{\rm Z, acc}\simeq  9.3 \,{\rm Myr}\,\,Z^{-1}\, \eta   \,\left(\frac{E_{\rm CRs}}{\rm 10^{17}\,eV}\right)\,\left(\frac{B}{\rm 1\,\mu G}\right)^{-1}\,\left(\frac{\beta}{0.2}\right)^{-2}\,,    
	\ee
}
where $\eta\gtrsim$ 1  is the gyro-magnetic factor. Under assumption that UHECRs escape via diffusion, the timescale is
$t_{\rm esc}\simeq \frac{R^2}{2\,D(E)}$, where $D(E)=\frac13\eta R_g$ is the diffusion coefficient. Therefore, the  escape timescale  can be written as 
{\small
	\be\label{tdif}
	t_{\rm Z,esc}\simeq 400\,{\rm Myr} \,\,Z\,\eta^{-1}\,\left(\frac{R}{\rm 100\, kpc}\right)^2\,\left(\frac{B}{\rm 1\,\mu G}\right)\,\left(\frac{E_{\rm CRs}}{\rm 10^{17}\,eV}\right)^{-1} \,.
	\ee
}
Accelerated CRs unavoidable interact with the propagating medium via hadronuclear inelastic interaction. Therefore, the neutrino spectrum is related to the CR spectrum through
\be\label{Ap}
E_\nu L_\nu\simeq f_{Ap} E_A L_A\,,
\ee
where  $f_{Ap}=1-e^{-t_{esc}/t_{loss}}\simeq {\rm min} \{k_p\sigma_{Ap}\,n_p\, t_{\rm esc},\,1\}$ is efficiency of the process  $k_p\simeq$ 0.5, $\sigma_{Ap}\simeq 8\times 10^{-26} {\rm cm^2} \; A^{3/4}$ at $\sim$ 100 PeV with A the atomic weight, $n_p$ is the proton density of the medium and $t_{\rm esc}$ is the confinement timescale of CRs in a region. In this process the neutrino energy is related with cosmic energy E$_\nu \sim$ 0.05 E$_A$. Then, to produce a 2 PeV neutrino we require carbon nuclei with energies in the range $\sim 250-300$ PeV.
\section{Neutrino around of CenA}
Previously,  Cen A had been discarded as source of IceCube neutrino event by \cite{Saba2013}. They showed that neutrino flux produced in the core of CenA are very low to be detected by actual instruments like IceCube.  In addition,\cite{Fraija2014} analyzed the neutrino emission from the giant lobes via hadronuclear processes and found that the flux is also undetectable.\\
In this work we consider the possibility that CRs are accelerated by giant lobes of CenA and after they escape from acceleration region to be confined again by galaxies around of giant lobes in the CenA group.
Because of the strength of magnetic field and the gas density are higher than in the giant lobes,  these galaxies could enhance the photo-pion efficiency,  thus increasing the confinement time and reducing the loss time. \\
From Hillas criterium we can see that giant lobes could accelerate UHECRs up to energies $\sim$ 10$^{20}$ eV that are much higher to those energies necessary to produce a 2 PeV neutrino. Using $\eta \sim 1$ that lies in the Bohm limit and comparing the acceleration timescale and escape timescale with the age of giant lobes we found that $t_{acc} \lll t_{esc}$ and $t_{esc} \sim t_{lobes}$ for $\sim$ 300 PeV.  Therefore only CRs with energies larger than $\sim$ 250 PeV can escape from giant lobes and to be injected in Cen A intragroup medium (with ages of giant lobes between$\sim$ 500-600 Myr \cite{Wykes2013}). Therefore, it is hard to expect neutrinos with energies lower than $\sim$ 2 PeV in a region around the giant lobes.\\
\\ 
Considering the scenario where CRs are trapped by galaxies close to the giant lobes, we found two types of dwarf galaxies in the CenA group: dE-type and  dIrr-type.  dE-type (elliptical dwarf)  are formed with gas poor and dIrr-type (Irregular dwarf) are gas rich and one Spiral Starburst galaxy. They are shown in the table \ref{table4}.   We consider the galaxies located at distance $\sim$ 4 Mpc around the giant lobes and inside the angular error of IC35 neutrino event.
To calculate the photo-pion efficiency we need to know the gas density and the magnetic field of the galaxies. Following  \cite{Kennicutt1998} is possible to relate the gas contained and star formation rate in a galaxy
\be
\Sigma_{SFR}=(2.5\pm0.7)\times10^{-4} \left(\frac{\Sigma_{\rm gas}}{\rm M_\odot \rm pc^{-2}}\right)^{1.4\pm0.15} {\rm M_\odot yr^{-1} kpc^{-2}}.
\ee
This relation still seems to hold in the case of dwarf galaxies \cite{Chyzy2011}. IN the case of dIrr we assume that star formation occur inside a radius $\sim$ 1 kpc  and scale height disk of $\sim$ 500 pc. Then the gas density is
\be
n_g\simeq 24 \left(\frac{\rm SFR}{\rm M_\odot \, {\rm yr^{-1}}}\right)^{0.71}\, \left( \frac{\rm R}{{\rm kpc}} \right)^{-1.43}\, \left(\frac{l}{500 {\rm  pc}} \right)^{-1} {\rm cm^{-3}}
\ee
For the case of dE, we assume a similar lower limit of density gas \cite{Bouchard2007} due to this type of galaxies are gas poor and large amount of their gas had been stripped by intragroup medium .
\\
The other parameter necessary to calculate the diffusion timescale is the strength of the magnetic field which can be calculated following \cite{Thompson2007}
\begin{equation}
B \simeq 5.864 \; \mu G \; \left(\frac{n_g}{\rm cm^{-3}} \right)^{0.7} \left(\frac{l}{500 \, \rm pc} \right)^{0.7}
\end{equation}
In table \ref{table4}, it can be observed that the highest values of photo-pion efficiency are related with ESO324-G024 and NGC4945 which exhibit 0.865 and 1, respectively.
\begin{table}[ht!]\label{tb}
	\centering
	\caption{ Galaxies located at a distance of $\sim$ 4 Mpc around of giant lobes and inside a error circle of IC35 event. The column 1 shows the galaxy name, column 2 denote the type, column 3 the number density of gas, column 4 the strength of the magnetic field, column 5 the loss time, column 6 the diffusion time of CRs inside of galaxies and column 7 the photo-pion efficiency.}\label{table4}
	
	\begin{tabular}{ l c c c c c c c c}
		\hline
		Galaxy & \scriptsize{Type} 
		& \scriptsize{$n_{g}$} &\scriptsize{B}& \scriptsize{$t_{loss}$} & \scriptsize{$t_{esc}$} &\scriptsize{$f_{Ap}$} & \scriptsize{}\\
		
		& \scriptsize{}& 
		& \scriptsize{(cm$^{-3}$)} &\scriptsize{($\mu$G)}& \scriptsize{(Myear)}&\scriptsize{(Myear)}&\scriptsize{}& \scriptsize{}\\
		\hline 
		\hline\\
		
		\scriptsize{ESO324-G024}   & \scriptsize{dIrr} 
		 &  \scriptsize{15} &\scriptsize{39.9}&\scriptsize{3.2}& \scriptsize{4.5} & \scriptsize{0.865}&  \scriptsize{(1)} \\
		
		\scriptsize{ESO325-G011}   & \scriptsize{dIrr} 
		&  \scriptsize{1.39} &\scriptsize{7.6}&\scriptsize{34.5}& \scriptsize{0.3}& \scriptsize{$8.7 \times 10^{-3}$}& \scriptsize{(3)} \\		

		\scriptsize{NGC5237}   & \scriptsize{dIrr}  
		& \scriptsize{2.14} &\scriptsize{10.2}&\scriptsize{22.5}& \scriptsize{0.6}& \scriptsize{$2.6 \times 10^{-2}$}& \scriptsize{(3)} \\

		\scriptsize{AM1339-445}   & \scriptsize{dE} 
		&  \scriptsize{$<10^{-2}$} &\scriptsize{0.2}&\scriptsize{$4.7 \times 10^{3}$}& \scriptsize{$1.5 \times 10^{-2}$}& \scriptsize{$3.2 \times 10^{-6}$}& \scriptsize{(1)} \\

		\scriptsize{AM1343-452}   & \scriptsize{dE} 
		 &  \scriptsize{$<10^{-2}$} &\scriptsize{0.2}&\scriptsize{$4.7 \times 10^{3}$}& \scriptsize{$1.5 \times 10^{-2}$}& \scriptsize{$3.2 \times 10^{-6}$}& \scriptsize{(1)} \\
				
		\scriptsize{UKS1424-460}   & \scriptsize{dIrr} 
		 &  \scriptsize{0.18} &\scriptsize{1.8}&\scriptsize{$2.7 \times 10^{2}$}& \scriptsize{$1 \times 10^{-1}$}& \scriptsize{$3.7 \times 10^{-4}$}& \scriptsize{(3)} \\
		
		\scriptsize{ESO174-G001}   & \scriptsize{dE} 
		&  \scriptsize{$<10^{-2}$} &\scriptsize{0.2}&\scriptsize{$4.7 \times 10^{3}$}& \scriptsize{$1.5 \times 10^{-2}$}& \scriptsize{$3.2 \times 10^{-6}$}& \scriptsize{(1)}\\

		\scriptsize{SGC1319.1-426}   & \scriptsize{dE} 
		&  \scriptsize{$<10^{-2}$} & \scriptsize{0.2} & \scriptsize{$4.7 \times 10^{3}$}&\scriptsize{$1.5 \times 10^{-2}$} & \scriptsize{$3.2 \times 10^{-6}$}& \scriptsize{(1)} \\

		\scriptsize{ESO269-G066}   & \scriptsize{dE} 
		&  \scriptsize{$<10^{-2}$} & \scriptsize{0.2}&\scriptsize{$4.7 \times 10^{3}$} &\scriptsize{$1.5 \times 10^{-2}$} & \scriptsize{$3.2 \times 10^{-6}$}& \scriptsize{(1)} \\

		\scriptsize{ESO269-G058}   & \scriptsize{dIrr} 
		&  \scriptsize{0.26} &\scriptsize{2.3}&\scriptsize{$1.8 \times 10^{2}$} & \scriptsize{$1.3 \times 10^{-1}$}& \scriptsize{$7.2 \times 10^{-4}$}& \scriptsize{(3)} \\
								
		\scriptsize{NGC4945}   & \scriptsize{SB(s)cd} 
		 &  \scriptsize{140} &\scriptsize{$1.9 \times 10^{2}$}&\scriptsize{0.4}& \scriptsize{7.2}& \scriptsize{1}& \scriptsize{(2)} \\

		\scriptsize{ESO269-G037}   & \scriptsize{dIrr} 
		&  \scriptsize{$<10^{-2}$} &\scriptsize{0.2}&\scriptsize{$4.7 \times 10^{3}$}& \scriptsize{$1.5 \times 10^{-2}$}& \scriptsize{$3.2 \times 10^{-7}$}& \scriptsize{(1)}\\
		
		\hline
	\end{tabular}
	\begin{flushleft}
		\scriptsize{
			\textbf{References}.  
			$^{(1)}$ \cite{Bouchard2007}; $^{(2)}$ \cite{Cote2009}; 
			$^{(3)}$ \cite{Chou2007}    }
	\end{flushleft}

\end{table}
%
\section{Conclusion}
In this work we have analyzed the possible emission of the event IC35 from regions around of CenA.  In this scenario,  giant lobes of CenA provides a powerful place  to accelerate UHECRs at energies as high as EeV.  In the scenario where accelerated carbon nuclei escape from giant lobes and later are confined by galaxies, especially ESO324-G024 and NGC4945, we find that although nuclei-proton efficiencies are high in these places cannot produce a PeV neutrino event in IceCube detector. Therefore, we do not find enough evidence to correlate IC35 with UHECRs detected by PAO.
Details about the chemical composition, angle deflexions and acceleration of UHECRs, as well as PeV neutrino calculations can be found in \cite{Fraija2018}.

\section*{Acknowledgments}
This work was supported by UNAM PAPIIT grant IA102917.
\section*{References}


\smallskip

\end{document}